\documentclass[runningheads]{llncs}

\usepackage{graphicx}
\usepackage{subcaption}
\usepackage{soul}
\usepackage{enumitem}

\title{Addressing the Scalability Bottleneck of Semantic Technologies at Bosch}

\authorrunning{{}}
\titlerunning{{}}
\author{ 
Diego Rincon-Yanez\inst{1,2} \and
Mohamed H. Gad-Elrab\inst{1}\and
Daria Stepanova\inst{1} \and 
Kien Trung Tran\inst{1} \and 
Cuong Chu Xuan\inst{1}\and
Baifan Zhou\inst{3,4} \and
Evgeny Karlamov\inst{1,3}
}

\institute{
Bosch Center for AI, Germany \and
University of Salerno, Italy \and
SIRIUS Centre, University of Oslo, Norway \and
Oslo Metropolitan University, Norway
}

\begin{document}

\maketitle

\medskip\noindent 
\section{Introduction} At the heart of smart manufacturing is real-time semi-automatic decision-making. Such decisions are vital for optimizing production lines, e.g., reducing resource consumption, improving the quality of discrete manufacturing operations, and optimizing the actual products, e.g., optimizing the sampling rate for measuring product dimensions during production. Such decision-making relies on massive industrial data thus posing a real-time processing bottleneck.

Indeed, consider an example of automated welding that is present in multiple Bosch production sites where real-time decisions include welding machine adjustment when welding quality (welding spots) degrades~\cite{DBLP:journals/ws/ZhouSGSCMWK21}. 
Such processes are data-intensive, including sensor measurements, e.g.,  temperature, pressure, and electrical conductivity, settings of welding parameters, and replacement of accessories (welding caps), etc.  When the welding is performed during car body manufacturing, each such body has up to 6.000 welding spots, generating  a large amount of data instances. The data is distributed across several analytical pipelines in charge of feed statistics, training traditional ML models, quality control measures, and many more~\cite{Zheng2022c}. 

This decision-making for welding requires both integration of heterogeneous data and real-time computation on top of it, thus leading to scalability bottleneck. Delivering a consistent and accurate industry-grade solution as long as the data grows in time, increases the complexity in scalability and performance terms. These challenges combined with the need of maintaining daily operations, increase further scalability requirements, creating the need to implement additional machine learning, knowledge engineering, and data management solutions transversely into a unified framework.

Bosch is a multinational company with a strong emphasis on manufacturing and engineering in automotive, energy, consumer goods, and other industries. Smart and AI-powered manufacturing is one of the central pillars of the Bosch strategy,  thus real-time effective and efficient processing of extreme data chains of heterogeneous, distributed~\cite{Rincon-Yanez2021}, fast-growing, and often disconnected or hardly compatible information is critical for the company's success.

\section{Semantic Approach to Industrial-Scale Data}
\begin{figure}[t]
\vspace{-4ex}
\centering
       \includegraphics[width=1\linewidth]{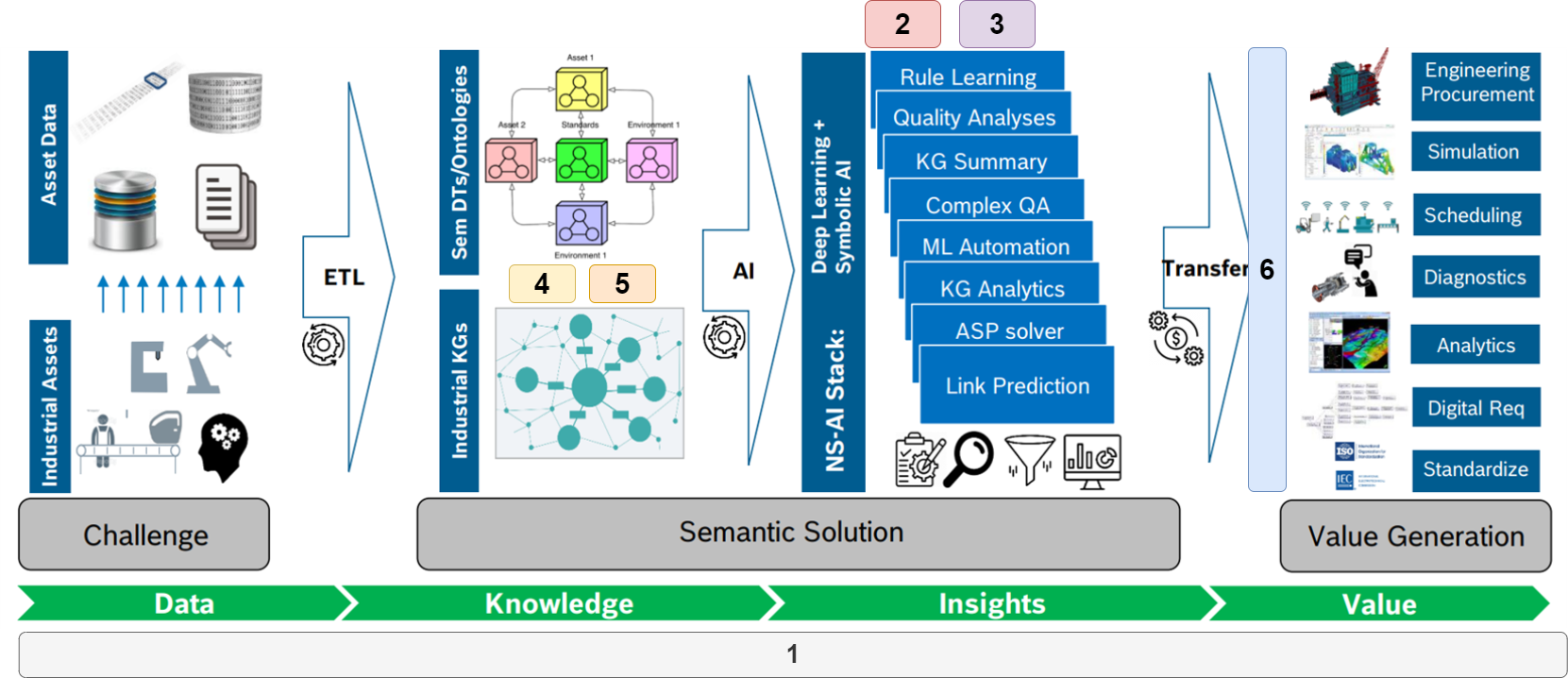}
       \vspace{-4ex}
       \caption{Semantic approach at Bosch to deal with large industrial data}
       \label{fig:roadmap} 
       \vspace{-4ex}
\end{figure}

In Bosch, we follow a semantic approach~\cite{Zheng2022}  to deal with large-scale industrial data, as depicted in Figure~\ref{fig:roadmap}.  
The idea is to unlock the value of data by exploiting industrial knowledge graphs to support decision-making. 
In particular, the data is first converted into KGs via ETL processes\cite{Rincon-Yanez2020}, then analyzed using Neuro-Symbolic AI methods that combine both semantic technology and Machine Learning, and finally, the results of analyses are transferred to industrial applications. This allows to bridge the data challenge and the value generation part of manufacturing. 

In order to ensure that the proposed approach offers the expected value, Bosch does a strong focus on both research and system development in scalability. 
In particular, Bosch does it via in-house research efforts and libraries and as a part of several EU projects such as enRichMyData on at-scale data annotation pipelines (\url{https://enrichmydata.eu/}), GraphMassivizer  on massive processing of graph data (\url{https://graph-massivizer.eu/}), DataCloud on scalable automated deployment of data pipelines to the Cloud (\url{https://datacloudproject.eu/}), and SmartEdge on edge-driven computations (\url{https://www.smart-edge.eu/}). 

\begin{figure}[t]
\vspace{-4ex}
\centering
       \includegraphics[width=0.6\linewidth]{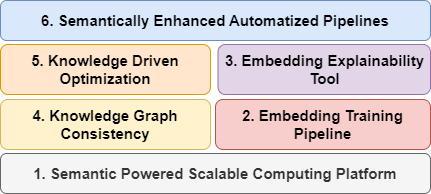}
       \vspace{-2ex}
       \caption{Stacked view of semantic solutions for Neuro-Symbolic use-cases at Bosch}
       \label{fig:stack} 
       \vspace{-2ex}
\end{figure}

Our scalable Neuro-Symbolic AI-powered ecosystem as can be seen in Figure \ref{fig:stack}, comprehend a set of tools, libraries, and frameworks destined to integrate, process, and deploy traditional data pipelines, and Industrial KG; these orchestrated components have the objective to empower the experienced and non-experienced internal users to leverage value from incoming from the data generated by the production lines in the different Bosch manufacturing scenarios.

The solution to address the scalability bottleneck covers the  transversely company semantic approach, a high-level view is given in Figure \ref{fig:roadmap}, following characteristics:

\begin{enumerate}[topsep=3pt,parsep=0pt,partopsep=0pt,itemsep=0pt,leftmargin=*]
    \item \textit{Semantic Powered Scalable Computing platform:} Consists of a scalable execution environment specially designed for simulating and executing data pipelines, this environment implements underlying semantic capabilities to assist in the pipeline description as well as exploiting the hardware configurations to find the best deployment scenario based on an initial requirement set.\looseness=-1
    \item \textit{Embedding Training Pipeline:} Provides a standardized pipeline for embeddings computation, merging symbolic reasoning with traditional low-dimension vector representations, delivering a fast prototyping experimentation tool seamlessly integrated with the internal data silos in the company environment. \looseness=-1  
    \item \textit{Embedding Explainability Tool:} Library with the capability of analyzing semantic enhanced KG with previously trained embedding models comparing different models providing understandability methods for knowledge representation tasks~\cite{gad2020excut,zheng2022towardsstats}.
    \item \textit{Knowledge Graph Consistency Check: } Tool to discover inconsistency patterns in KGs with respect to ontological rules, this allows to processing of the data as well as the recognition of inconsistency patterns in the evaluated KG~\cite{kien2020inconsistency}.
    \item \textit{Knowledge Driven Optimization:} Package for encoding generic KG-based optimization problems in Answer Set Programming (ASP) language, which is currently used in suppliers optimization, factory planning, and scheduling.
    \item \textit{Semantically Enhanced Automatized pipelines: } In order to deliver efficiency to the internal teams template-typed pipelines were prepared to leverage the use of mappings techniques to create KB's to exploit traditional ML knowledge~\cite{zheng2022executable,klironomos2023exekglib}.
\end{enumerate}

\section{Conclusion}

Given the specific nature of the use case and the particular needs of a company like Bosch, there is no off-the-shelf software solutions able to adapt the constantly grow and demands in terms of data streaming speed, security and volume.
As lessons learned, we deeply understand that semantic technology not offers shared data schema that unifies different data syntax and semantics but also 
offers unambiguous ``lingua franca'' for cross-domain communication, that unifies the language and understanding of stakeholders~\cite{yahya2023semantic}. This greatly helps the stakeholders to perform tasks of a remote domain (e.g., semantic technology) that otherwise would be error-prone, time-consuming and cognitively demanding.
Meanwhile, it is important to be flexible in adopting combined technology of relational database and KG, such as a mixture of both or virtual KG, to exploit the flexibility of KG as well as the computational performance of relational databases.\looseness=-1

Modern automatic manufacturing requires real-time decision-making for quality optimization tasks. ML methods face new challenges and opportunities to holistically analyze the massive and unprecedented data integrated across these chains, In this tailored ecosystem the goal is to speed up the experimentation capabilities of the onsite teams by providing a set of state-of-the-art semantically enhanced evaluation approaches, to support decisions that  change the manufacturing processes towards a sustainable, circular, and climate-neutral industry. \looseness=-1

Finally, we are excited to present our Bosch challenges and solutions to address the scalability bottleneck for semantic data to the ESWC community. We believe our case of large industrial data is rather typical for large manufacturing and service industries and thus will be of interest to a wide audience. 

\smallskip
\noindent \textbf{Acknowledgements}:
The work was partially supported by EU projects: Dome 4.0 (GA 953163), OntoCommons (GA 958371), DataCloud (GA 101016835), Graph Massiviser (GA 101093202) and enRichMyData (GA 101093202), SMARTEDGE (GA
101092908) and the SIRIUS Centre, (NRC, No. 237898).

\bibliographystyle{splncs04}
\bibliography{biblio.bib}

\begin{thebibliography}{10}
\providecommand{\url}[1]{\texttt{#1}}
\providecommand{\urlprefix}{URL }
\providecommand{\doi}[1]{https://doi.org/#1}

\bibitem{gad2020excut}
Gad-Elrab, M.H., Stepanova, D., Tran, T.K., Adel, H., Weikum, G.: Excut:
  Explainable embedding-based clustering over knowledge graphs. In: ISWC 2020.
  pp. 218--237. Springer (2020)

\bibitem{klironomos2023exekglib}
Klironomos, A., Zhou, B., Tan, Z., Zheng, Z., Mohamed, G.E., Paulheim, H.,
  Kharlamov, E.: {ExeKGLib}: Knowledge graphs-empowered machine learning
  analytics. In: ESWC (Demo \& Posters) (2023)

\bibitem{Rincon-Yanez2021}
Rincon-Yanez, D., Crispoldi, F., Onorati, D., Ulpiani, P., Fenza, G., Senatore,
  S.: {Enabling a Semantic Sensor Knowledge Approach for Quality Control
  Support in Cleanrooms}. In: ISWC. vol.~2980 (2021)

\bibitem{Rincon-Yanez2020}
Rincon-Yanez, D., Lauro, E.D., Falanga, M., Senatore, S., Petrosino, S.:
  {Towards a semantic model for IoT-based seismic event detection and
  classification}. In: 2020 IEEE SSCI. pp. 189--196. IEEE (dec 2020)

\bibitem{kien2020inconsistency}
Tran, T.K., Gad-Elrab, M.H., Stepanova, D., Kharlamov, E., Str\"{o}tgen, J.:
  Fast computation of explanations for inconsistency in large-scale knowledge
  graphs. In: Proceedings of The Web Conference 2020. p. 2613–2619. WWW
  (2020)

\bibitem{yahya2023semantic}
Yahya, M., Zhou, B., Breslin, J.G., Ali, M.I., Kharlamov, E.: Semantic
  modeling, development and evaluation for the resistance spot welding
  industry. IEEE Access  (2023)

\bibitem{zheng2022towardsstats}
Zheng, Z., Zhou, B., Zhou, D., Khan, A.Q., Soylu, A., Kharlamov, E.: Towards a
  statistic ontology for data analysis in smart manufacturing. In: ISWC (Demo
  \& Posters) (2022)

\bibitem{Zheng2022c}
Zheng, Z., Zhou, B., Zhou, D., Soylu, A., Kharlamov, E.: {Executable Knowledge
  Graph for Transparent Machine Learning in Welding Monitoring at Bosch}. In:
  CIKM. pp. 5102--5103 (2022)

\bibitem{zheng2022executable}
Zheng, Z., Zhou, B., Zhou, D., Soylu, A., Kharlamov, E.: Executable knowledge
  graph for transparent machine learning in welding monitoring at {Bosch}. In:
  Proceedings of the 31st ACM International Conference on Information \&
  Knowledge Management. pp. 5102--5103 (2022)

\bibitem{Zheng2022}
Zheng, Z., Zhou, B., Zhou, D., Zheng, X., Cheng, G., Soylu, A., Kharlamov, E.:
  {Executable Knowledge Graphs for Machine Learning: A Bosch Case of Welding
  Monitoring}. In: LNCS, vol. 13489 LNCS, pp. 791--809. Springer (2022)

\bibitem{DBLP:journals/ws/ZhouSGSCMWK21}
Zhou, B., Svetashova, Y., Gusmao, A., Soylu, A., Cheng, G., Mikut, R., Waaler,
  A., Kharlamov, E.: Semml: Facilitating development of {ML} models for
  condition monitoring with semantics. J. Web Semant.  \textbf{71},  100664
  (2021)

\end{thebibliography}

\end{document}